\documentclass[aps,prl,twocolumn,floatfix,nofootinbib,showpacs]{revtex4}
\usepackage{graphicx}
\usepackage{color}

\newcommand{\U}{{\cal U}}
\begin{document}

%\begin{flushright}}
%\preprint{hep-ph/yymmnnn}
%\end{flushright}

%\bibliographystyle{revtex}

\title{Collider Signals of Unparticle Physics}

\renewcommand{\thefootnote}{\fnsymbol{footnote}}

\author{ 
Kingman Cheung$^{1,2}$, 
Wai-Yee Keung$^3$
and Tzu-Chiang Yuan$^1$
 }
\affiliation{$^1$Department of Physics, National Tsing Hua University, 
Hsinchu 300, Taiwan
\\
$^2$Physics Division, National Center for Theoretical Sciences,
Hsinchu 300, Taiwan
\\
$^3$Department of Physics, University of Illinois, Chicago Illinois 60628, USA
\\
{\rm Published in Phys. Rev. Lett. {\bf 99}, 051803 (2007)}
}

\renewcommand{\thefootnote}{\arabic{footnote}}
%\date{}

\begin{abstract}
Phenomenology of the notion of an unparticle $\U$, recently perceived
by Georgi, to describe a scale invariant sector with a non-trivial
infrared fixed point at a higher energy scale is explored in details.
Behaving like a collection of $d_\U$ (the scale dimension of the
unparticle operator $\cal O_U$) invisible massless particles, this
unparticle can be unveiled by measurements of various energy
distributions for the processes $Z \to f \bar f \U$ and $e^- e^+ \to
\gamma \U$ at $e^-e^+$ colliders, as well as monojet production at
hadron colliders.  We also study the propagator effects of the
unparticle through the Drell-Yan tree level process and the one-loop
muon anomaly.

\end{abstract}


\maketitle
 
{\it Introduction.}  
Scale invariance is a powerful concept that has
wide applications in many different disciplines of physics. 
In phase transitions and critical phenomena, the system becomes 
scale invariant at critical temperature since fluctuations at all
length scales are
important.  
In particle physics, scale invariance has also been a
powerful tool to analyze asymptotic behaviors of correlation
functions
at high energies. 
In string theory, scale invariance plays an even more
fundamental role since it is part of the local diffeomorphism $\times$ Weyl
reparametrization invariance group of the 2-dimensional Riemann
surfaces.  However, at the low energy world of particle
physics, what we observe is a plethora of elementary and composite
particles with a wide spectrum of masses \cite{pdg}. Scale invariance
is manifestly broken
by the masses of these particles.  Nevertheless, it is conceivable
that
at a much higher scale, beyond the Standard Model (SM), there is a
nontrivial scale invariant sector with an infrared fixed point that
we have not yet probed experimentally. For example, this sector can be
described by the vector-like non-abelian gauge theory with a large
number of massless fermions as studied by Banks and Zaks
\cite{banks-zaks}.

Recently, Georgi \cite{georgi} made an interesting
observation that a nontrivial scale invariance sector of scale
dimension $d_{\cal U}$ might manifest itself at low energy as a
non-integral number $d_{\cal U}$ of invisible massless particles,
dubbed unparticle $\U$.
%%%
It may give  rise to peculiar missing energy distributions at various
processes that can be probed at Large Hadron Collider (LHC) or
$e^- e^+$ colliders.  In this Letter, we explore
in details various implications of the unparticle $\U$ using the
language of effective field theory as in \cite{georgi}.  
We show that the energy distributions for the processes of $Z \to f \bar f \U$
at LEP and monophoton production plus missing energy via $e^-e^+ \to
\gamma \U$ at LEP2
can discriminate the scale dimension $d_\U$ of the unparticle, while
monojet production plus missing energy at the LHC
cannot easily do so because of parton smearing.
In addition, we generalize the
notion of real unparticle emission to off-shell exchange and study its
propagator effects in the Drell-Yan tree-level process and the muon
anomaly at one-loop level. We show that the invariant mass spectrum of
the
lepton pair in Drell-Yan process can discriminate the scale dimension
$d_\U$,
and
we can use the muon anomalous magnetic moment data to constrain the
scale dimension  as well as the effective coupling.

{\it Unparticle.}  For definiteness we denote the scale invariant
sector as a Banks-Zaks ($\cal BZ$) sector \cite{banks-zaks} and follow
closely the scenario studied in \cite{georgi}.  The $\cal BZ$ sector
can interact with the SM fields through the exchange of a connector
sector that has a high mass scale $M_\U$. Below this high mass scale,
non-renormalizable operators that are suppressed by inverse powers of
$M_\U$ are induced. Generically, we have operators of the form
\begin{equation}
   {\cal O}_{\rm SM}  \, {\cal O}_{\cal BZ} /{M_\U^k} \, \;
  \;\;\;\; (k > 0) \;,
\label{genericop}
\end{equation}
where ${\cal O}_{SM}$ and ${\cal O_{BZ}}$ represent local operators
constructed out of SM and ${\cal BZ}$ fields, respectively. As in
massless non-abelian gauge theories, renormalization effects in the
scale invariant $\cal BZ$ sector induce dimensional transmutation
\cite{coleman-weinberg} at an energy scale $\Lambda_\U$ . Below
$\Lambda_\U$ matching conditions must be imposed onto the operator
(\ref{genericop}) to match a new set of operators having the following
form
\begin{equation}
 \label{effectiveop}
   (C_{\cal O_U} \Lambda^{d_{\cal BZ} - d_\U} / {M^k_\U } ) \, 
   {\cal   O}_{\rm SM}\,   {\cal O}_\U \;,
\end{equation}
where $d_{\cal BZ}$ and $d_\U$  are the scale dimensions of  
$\cal O_{BZ}$ and the unparticle operator ${\cal O_U}$ respectively,
and
$C_{\cal O_\U}$ is a coefficient function fixed by the matching.

Three unparticle operators with  different Lorentz structures were
addressed in \cite{georgi}:
 $\left\{ O_\U, O^\mu_\U, O_\U^{\mu\nu} \right\} \in {\cal O_U}$. It
was argued in \cite{georgi} that 
 scale invariance can be used to fix the two-point functions of these
 unparticle operators. For instance,
 \begin{equation}
  \langle 0 \vert O_\U (x)  O_\U^\dagger (0) \vert 0 \rangle =
 \int \frac{ d^4 P}{(2\pi)^4}   
e^{-i P \cdot x} \, \vert \langle 0 \vert O_\U (0)\vert P
 \rangle \vert^2
\rho(P^2) \, 
\end{equation}
with
$\vert \langle 0 \vert O_\U (0)\vert P \rangle \vert^2
\rho(P^2)  \, = \,
 A_{d_\U}\, \theta(P^0) \, \theta (P^2) \,(P^2)^{d_\U -2 } 
$
where $A_{d_\U}$ is normalized to interpolate 
the $d_\U$-body phase space of  massless
particle \cite{georgi} 
%%%%
%
\begin{equation}
   A_{d_\U}={16\pi^2\sqrt{\pi}\over (2\pi)^{2{d_\U}}}
       { \Gamma({d_\U}+{1\over
       2})\over\Gamma({d_\U}-1)\Gamma(2\,{d_\U})}  \; .
\end{equation}
These unparticle operators are all taken to be Hermitian, and
$O^\mu_\U$ and $O^{\mu\nu}_\U$ are assumed to be transverse. 
As pointed out in \cite{georgi}, important effective operators of the
form (\ref{effectiveop}) that can give rise to interesting
phenomenology are
\begin{eqnarray}
\lambda_0 \frac{1}{\Lambda_\U^{d_\U} } G_{\alpha\beta} G^{\alpha\beta}
O_\U \;\; & , &  \;\;
\lambda_1 \frac{1}{\Lambda_\U^{d_\U - 1} }\, \bar f \gamma_\mu f \,
O_\U^\mu \nonumber \\
\lambda_2 \frac{1}{\Lambda_\U^{d_\U} } G_{\mu\alpha}
G_{\nu}^{\;\alpha} O_\U^{\mu\nu} 
\;\; & , & \;\; {\rm etc.,}
\label{uops}
\end{eqnarray}
where $G^{\alpha\beta}$ denotes the gluon field strength, $f$ stands
for a SM fermion, and $\lambda_{i}$ are dimensionless effective
couplings $C_{O^i_\U} \Lambda_\U^{d_{\cal BZ}}/M_\U^k$ with the index
$i=0,1$ and $2$ labeling the scalar, vector and tensor unparticle
operators respectively.  The scalar operator $O_\U$ can also couple to
the SM fermions. However, its effect is necessarily suppressed by the
fermion mass. We focus on the first two operators of
Eq.(\ref{uops}) in this work. 
For simplicity, we assume universality
that $\lambda_1$ is flavor blind. Furthermore, we only consider 
$d_{\cal  U} \ge 1$ to avoid the crash with unitarity of the theory \cite{mack}.

{\it Phenomenology.} 
We now turn to several phenomenological implications of the
unparticle.

\begin{figure}[t!]
\includegraphics[width=3.3in]{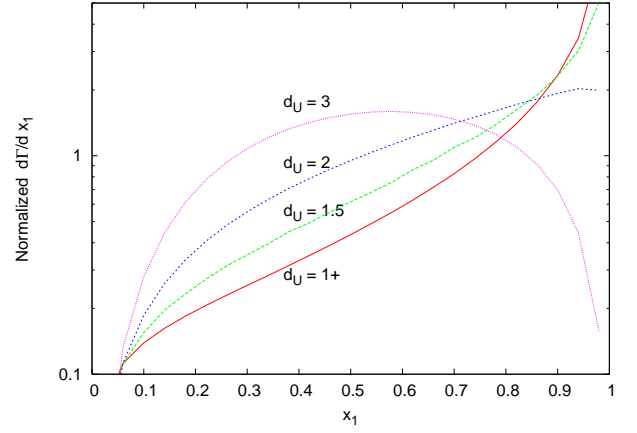}
\caption{
\label{zqqu} \small
Normalized decay rate of $Z \to q \bar q {\cal U}$  versus $x_1 = 2
E_f/M_Z$ for different values of $d_{\cal U}=1+\epsilon,\,1.5,\,2$, 
and 3 with $\epsilon$ a small number.
}
\end{figure}

\noindent
{\it (1) $Z \to f \bar f \U$}: The decay width for the process can be
easily obtained as 
\begin{eqnarray}
\label{ztoqqU}
\frac{1}{\Gamma_{Z \to f \bar f} }
\frac{d\Gamma(Z \to f \bar f +{\cal U}) }
{ d x_1 d x_2 d \xi} & =
 &
 \frac{\lambda_1^2}{8\pi^3} \, g(1-x_1,1-x_2,\xi) \nonumber \\
& \times & \frac{M_Z^2} {\Lambda_\U^2}
 A_{d_{\cal U}} \left(\frac{P_\U^2}{\Lambda_\U^2}\right)^{d_{\cal
 U}-2}
\end{eqnarray}
where $ \xi = P_\U^2/M_Z^2$ and $x_{1,2}$ are the energy fractions of
the fermions $x_{1,2} = 2 E_{f, \bar f} / M_Z$. The function
$g(z,w,\xi)$ is given by
\begin{eqnarray}
g(z,w,\xi) &=&
\frac{1}{2}\, \left( \frac{w}{z} +\frac{z}{w} \right)
+ \frac{(1+\xi)^2}{zw} \nonumber \\
  &-& \frac{\xi}{2}\left(\frac{1}{z^2}+\frac{1}{w^2}
\right)  
    -    \frac{1+\xi}{z} - \frac{1+\xi}{w} \; .
\end{eqnarray}
The integration domain for Eq.(\ref{ztoqqU}) is defined by $0 < \xi <
1, 0 < x_1 < 1 - \xi$ and $1- x_1 - \xi < x_2 < (1 - x_1 - \xi )/( 1 -
x_1)$.  In Fig.~\ref{zqqu}, we plot the normalized decay rate of this
process versus the energy fraction of the fermion $f$.  One can see
that the shape depends sensitively on the scale dimension of the
unparticle operator. As $d_\U \to 1$, the result approaches to a
familiar case of $\gamma^* \to q \bar q g^*$
\cite{hagiwara-martin-stirling}.  

\begin{figure}[t!]
\includegraphics[width=3.3in]{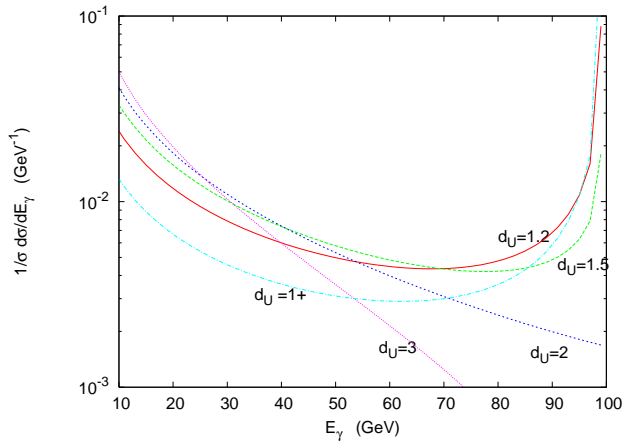}
\caption{
\label{eegammau} \small
Normalized monophoton energy spectrum of $e^- e^+ \to \gamma {\cal
  U}$ 
for $d_{\cal U}=1+\epsilon,\,1.2,\,1.5,\,2$ and 3 at $\sqrt{s} = 200$ GeV.
We have imposed $|\cos\theta_\gamma|<0.95$.
}
\end{figure}

\noindent
{\it (2) Monophoton events in $e^- e^+$ collisions}: The energy
spectrum of
the monophoton from the process $e^- (p_1) \; e^+(p_2)\; \to \gamma
(k_1) \; \U (P_\U)$ can also be used to probe the unparticle.  Its
cross section is given by
\begin{equation} 
 d \sigma = \frac{1}{2 s} \,  | \overline{{\cal M}} |^2 \;
 \frac{  A_{d_\U} }{ 16 \pi^3 \Lambda_\U^2} 
 \left( {P^2_\U\over \Lambda_{\cal U}^2} \right )^{ d_\U - 2}  
\, E_\gamma d E_\gamma d \Omega  
\end{equation}
with the matrix element squared 
\begin{equation}
  |\overline{{\cal M}}|^2 = 2 e^2 Q_e^2 \lambda_1^2 \,
  \frac{ u^2 + t^2 + 2 s P^2_\U}{ u t} \; .
  \end{equation}
The $P^2_\U$ is related to the energy of the photon $E_\gamma$ by the
recoil mass relation, 
\begin{equation}  P^2_\U = s - 2 \sqrt{s} \, E_\gamma \ . 
\label{kinematic}
\end{equation}
The monophoton energy distribution is
plotted in Fig.~\ref{eegammau} for various choices of $d_\U$. The
sensitivity of the scale dimension to the energy distribution can be
easily discerned.  
Monophoton events have been searched quite extensively at LEP
experiments
\cite{LEP} in some other contexts. 
Details of comparison with the data
and background analysis will be given in a forthcoming publication
\cite{next}.

\noindent
{\it (3) Monojet at hadronic collisions:} It was suggested in
\cite{georgi} that at the hadronic collider, the following partonic
subprocesses
\begin{eqnarray}
 g      g  & \to &  g \U   \;,\; \;    q \bar q   \to  g \U     \;
 ,\nonumber \\ 
 q      g  & \to & q \U  \;,\; \; \bar q g \, \to \, \bar q \U
 \; \nonumber
\end{eqnarray}
which can lead to monojet signals could be important for
the detection of the unparticle.
For the subprocesses that involve both quark and gluon,
we consider solely the effects from the vector operator
$O^\mu_\U$. 
For the gluon-gluon fusion subprocess, we consider solely
the effects from the scalar operator $O_\U$. 
%
%%%%
Although $P_\U^2$ is related to $\hat s$ by a kinematic relation
similar to Eq.(\ref{kinematic}), 
it is not uniquely determined at the hadronic level
where $\hat s \sim x_1 x_2 s$ with $s$ the center-of-mass energy
squared of the colliding hadrons and $x_{1,2}$ are the parton momentum
fractions.  
We have studied in
details the $P^2_\U$ distribution in hadronic collisions.
We found that the peculiar feature of the phase space of fractional
$d_\U$ at partonic level is completely
washed out.  Therefore it would be difficult to detect the unparticle
at hadronic environment using the monojet signal, in contrast to
its original anticipation \cite{georgi}. Details will be presented elsewhere
\cite{next}.
 
\begin{figure}[t!]
\includegraphics[width=3.3in]{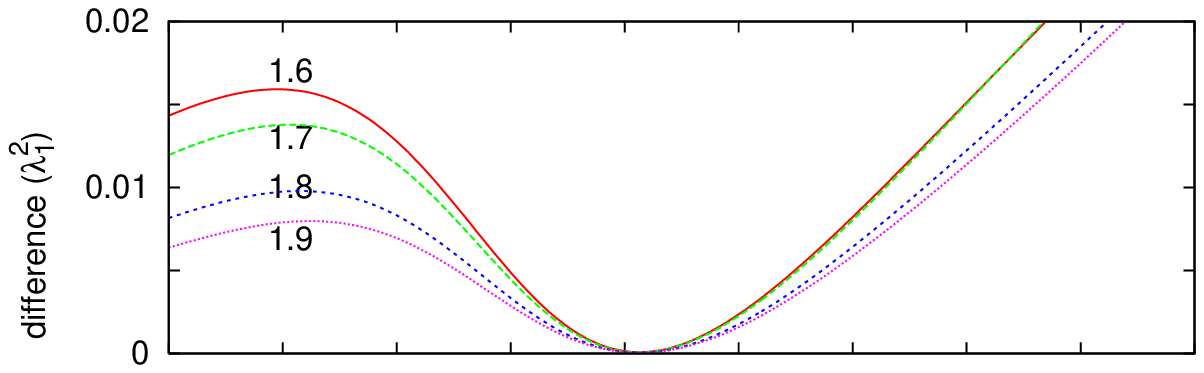}

\vspace{-0.29in}

\includegraphics[width=3.3in]{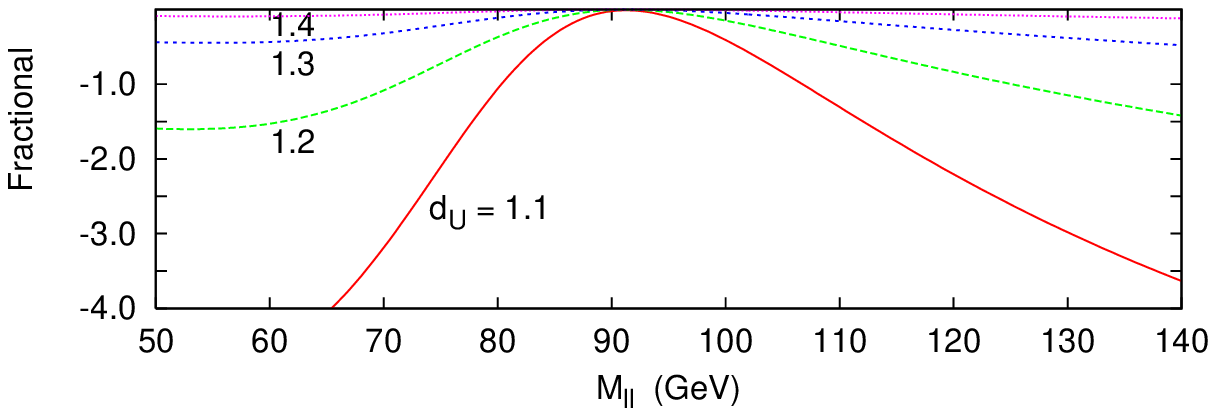}
\caption{
\label{drell-yan} \small
Fractional difference from the SM prediction of the Drell-Yan
invariant mass spectrum for various $d_{\cal U}$ at the Tevatron in
units of $\lambda_1^2$.  We have chosen $\Lambda_\U = 1$ TeV.  Note
that the scales in $\pm y$-axis are different. The curve for
$d_\U=1.5$ is too close to zero for visibility in the current scale.
}
\end{figure}

\noindent
{\it (4) Drell-Yan process:}
Using the K$\ddot{\rm a}$llen-Lehmann spectral representation formula,
the propagator for the vector unparticle operator $O_\U^\mu$ can be derived as
\begin{eqnarray}
\label{unpropagator}
\Delta_{F}^{\mu\nu}(P_\U^2)& =& Z_{d_\U} \left( -g^{\mu\nu} + 
  \frac{ P_{\U}^{\mu} P_{\U}^{\nu}} { P_\U^2} \right) 
\left( -P_\U^2 \right)^{d_\U -2}
\end{eqnarray}
with
\begin{equation}
Z_{d_\U} = \frac{A_{d_\U} }{2 \sin  (d_\U  \pi)} \;  \ .
\end{equation}
The $(-)$ sign in front of $P_{\cal U}^2$ of 
the unparticle propagator in  Eq.(\ref{unpropagator})
gives rise to a phase factor $e^{-i\pi d_{\cal U}}$ for  time-like
momentum  $P_{\cal U}^2>0$, 
but not for   space-like momentum $P_{\cal U}^2<0$.  
Virtual exchange of vector unparticle can result in the following 4-fermion interaction
\begin{eqnarray}
{\cal M}^{4f}_\U = \lambda_1^2 \, Z_{d_\U} 
\, \frac{1}{\Lambda_\U^2} \,
 \left(
- \frac{P_\U^2}{\Lambda_\U^2} 
  \right)^{d_\U - 2} \, 
( \bar f_1 \gamma_\mu f_2)\,  ( \bar f_3 \gamma^\mu f_4) 
\label{4fermionsop}
\end{eqnarray}
where the contribution from the longitudinal piece $P_\U^\mu
P_\U^\nu/P_\U^2$ 
has been dropped for massless external fermions.
Note that $P_\U^2$ is taken as the $\hat s$ for an $s$ channel
exchange  subprocess.  
The most important feature is that the high
energy
behavior of the amplitude scales as $(\hat s/\Lambda_\U^2 )^{d_\U  -1}$.  
For $d_\U=1$ the tree amplitude behaves like that of a massless
photon exchange, 
while for $d_\U = 2$ the amplitude reduces to the
conventional 4-fermion interaction \cite{eichten-lane-peskin}, 
i.e., its high-energy behavior
scales like
$s/\Lambda_\U^2$.  If $d_\U$ is between 1 and 2, say $3/2$, the
amplitude has the unusual behavior of $\sqrt{\hat s}/\Lambda_\U$ at
high energy.  
If $d_\U = 3$ the amplitude's high energy behavior becomes $(\hat
s/\Lambda_\U^2)^2$,
which resembles the exchange of Kaluza-Klein tower of gravitons
\cite{km}. But for virtual integration, one must restrict $d_{\cal U}<2$.
We can determine the differential cross section for the Drell-Yan
process
\begin{eqnarray}
&& \frac{d^2 \sigma}{ d M_{\ell\ell} \, dy}  =  K
  \frac{M^3_{\ell\ell}}{72\pi s}\,
 \sum_q \, f_q(x_1) f_{\bar q}(x_2) \;\;\;\;\;\;\;\;\;\; \nonumber \\
 && \times \left (
  | M_{LL} |^2 +   | M_{LR} |^2  +  | M_{RL} |^2  +  | M_{RR} |^2 
             \right ) \,,
\end{eqnarray}
where $\hat s = M^2_{\ell\ell}$ and $\sqrt{s}$ is the center-of-mass
energy of the colliding hadrons.  $M_{\ell\ell}$ and $y$ are
the invariant mass and the rapidity of the lepton pair, respectively,
and $x_{1,2} = M_{\ell\ell}e^{\pm y}/\sqrt{s}$.
The $K$ factor equals $1 + \frac{\alpha_s}{2\pi} \frac{4}{3} \left(
  1+ \frac{4 \pi^2}{3} \right )$.
The reduced amplitude $M_{\alpha\beta} (\alpha,\beta = L,R)$ is given
by
\begin{eqnarray}
 M_{\alpha \beta}  =  
 & \  &   \lambda_1^2 Z_{d_\U}  \frac{1}{\Lambda_\U^2} 
 \left (- \frac{\hat s}{\Lambda_\U^2} \right)^{d_\U-2} 
+ \frac{ e^2 Q_l Q_q}{ \hat s}    \nonumber\\
  &+& \frac{e^2 g^l_\alpha g^q_\beta}{ \sin^2 \theta_{\rm w}
  \cos^2\theta_{\rm w}
 }\, \frac{1}{\hat s - M_Z^2+iM_Z\Gamma_Z} 
 \label{reducedamp}
\end{eqnarray}
where $g_L^f = T_{3f} - Q_f \sin^2 \theta_{\rm w}$, $g_R^f = - Q_f
\sin^2 \theta_{\rm w}$ and $Q_f$ is the electric charge of the fermion
$f$. 
The phase $\exp (-i \pi d_\U )$ in the 4-fermion contact term 
will interfere with the $Z$ boson propagator 
in a rather non-trivial way.
This is because both the contact term phase and the $Z$ boson propagator
have the real and imaginary parts, which give rise to interesting
interference patterns \cite{Georgi:2007si}.  
This kind of interference had been studied 
some time ago in \cite{eichten-lane-peskin} 
in the context of preon models.
In Fig.~\ref{drell-yan}, we depict the 
fractional difference from the SM prediction 
in units of $\lambda_1^2$ (with small $\lambda_1$ 
while keeping $\Lambda_\U = 1$ TeV) of the Drell-Yan distribution as
a function of the invariant mass of the lepton pair for various 
$d_\U$. Interesting interference patterns around the $Z$ pole are easily 
discerned.
We can also allow different
couplings to different chirality combinations in the 4-fermion contact
interactions, denoted by $LL,RR,LR$ and $RL$.  For example, the combinations
of $LL+RR+LR+RL$ and $LL+RR-LR-RL$ give the $VV$ and $AA$ interactions 
respectively and these were studied in Ref. \cite{Georgi:2007si}.   
By doing so we can
reproduce the effects in Ref. \cite{Georgi:2007si}.
However, it may be difficult to
disentangle the fractional difference from the SM prediction
in Drell-Yan production due to experimental uncertainties.  It may be
easier to test the angular distributions and interference patterns in
$e^+ e^-$ collisions that we will delay to a full publication \cite{next}.

\begin{figure}[t!]
\includegraphics[width=3.3in]{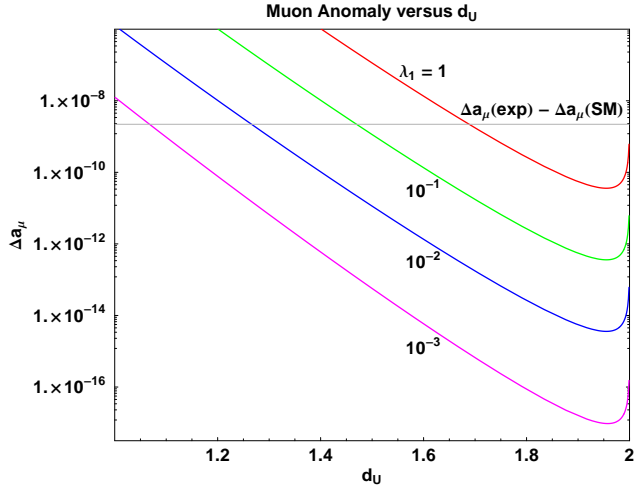}
\caption{
\label{muonanomaly} \small
Contribution to the muon anomalous magnetic moment from the unparticle 
versus $d_{\cal U}$  with
$\Lambda_\U = 1$ TeV and the coupling 
$\lambda_1 = 1,10^{-1},10^{-2}$ and 
$10^{-3}$.
}
\end{figure}

\noindent
{\it (5) Lepton anomalous magnetic moments:} Replacing one photon
exchange in QED by the unparticle associated with the vector operator
$O_\U^\mu$, one can derive the unparticle contribution to the lepton
anomaly $\Delta a_l=(g_l-2)/2$,
\begin{equation}
\Delta a_l = -\frac{\lambda_1^2 Z_{d_\U}}{4 \pi^2}   \left(
\frac{m^2_l}{\Lambda_\U^2}\right)^{d_{\U} - 1} 
{\Gamma (3 - d_{\U})\Gamma( 2 d_{\U} - 1) \over \Gamma(2+d_{\U})} \ ,
\end{equation}
where $m_l$ is the charged lepton mass.  
As $d_\U \to 1$, one has $\Delta a_l \to  \lambda^2_1 / 8 \pi^2$.
Setting $\lambda_1$ equals to  $e$, one reproduces the well
known QED result.  Note that the phase of the unparticle propagator 
does not appear in the lepton anomaly. 
A Wick rotation has effectively turned the
loop integral into spacelike and no phase can be picked up.
In Fig.~\ref{muonanomaly}, we plot the muon anomalous magnetic moment 
contribution from the unparticle
versus the scale
dimension $d_\U$ for various 
$\lambda_1$'s. 
The horizontal line is the experimental value of the
muon
anomalous magnetic moment with the SM contribution subtracted
\cite{pdg},
\begin{equation}
\Delta a_\mu ({\rm exp}) - \Delta a_\mu ({\rm SM}) = 22 (10) \times
10^{-10} \; .
\end{equation}
It is amusing to see that current experimental data of
the muon anomaly 
can give bounds to the effective coupling $\lambda_1$ and scale
dimension $d_\U$ already. 

{\it Conclusion.} Unparticle physics associated with a hidden scale
invariant sector with a nontrivial infrared
fixed point at a higher energy scale  has 
interesting phenomenological consequences at low
energy experiments. Effective field theory can be used to explore the
unparticle effects. Because the scale dimensions of the unparticle
operators can take on non-integral values, this leads to peculiar
features in the energy distributions for many processes involving SM
particles.  In this Letter, we have demonstrated these interesting
features can be easily exhibited for various processes in $e^- e^+$
machines, but not for the monojet production at hadron colliders
like the LHC. Moreover, virtual effects of the unparticle
could be seen in the Drell-Yan process and the muon anomaly.

{\it Acknowledgments.}
We thank M. Stephanov for a useful discussion.
This research was supported in parts by the NSC
under Grant No.\ NSC 95-2112-M-007-001, by the NCTS,
and by U.S. DOE under Grant No. DE-FG02-84ER40173.
We would like to thank Professor H. Georgi for comments on the 
manuscript and pointing out 
Refs.\cite{mack} and \cite{eichten-lane-peskin} to us. 

{\it Note added:} 
Recently, a second unparticle paper by Georgi appeared \cite{Georgi:2007si}, which
also studied the effect of the virtual propagation of the unparticle.   Our form of the 
unparticle propagator agrees with his, once we adopt the 
same normalization. 

%%%%%%%%%%%%%%%%%%%%%%%%%%%%%%%%%%%%%%%%

\end{document}